# Low-temperature differential-thermal analysis to measure variations in entropy

*A. Schilling and M. Reibelt*

*Physik-Institut der Universität Zürich, Winterthurerstrasse 190, CH-8057 Zürich*

**Abstract**

We describe how we can precisely measure variations in the entropy $S$ of small solid samples below room temperature, as a function the temperature $T$ or the external magnetic field $H$, respectively. A simple differential-thermal analysis (DTA) technique allows, in principle, for the measurement of variations in $S$ without any instrumental broadening of the data. The method is particularly well suited for the detection of sharp phase transitions in samples of milligram size.

## I. Introduction

Differential-thermal analysis (DTA) is widely used in chemical and material sciences for investigating the thermodynamics of chemical reactions and phase transitions[1]. In most cases, DTA measurements are used to qualitatively detect abrupt changes in the thermodynamic properties of a sample above room temperature. We have earlier extended this method to measure heat capacities of small samples below room temperature[2], and observed, as a result, novel electronic phase transitions in superconductors[3,4,5,6]. In this work we show that the same method can be used to directly measure variations in entropy $S$ as a function of the temperature $T$ or the magnetic field $H$, respectively. In contrast to corresponding heat-capacity $C$ data obtained by the same technique, variations in $S$ can be measured without introducing any instrumental broadening to the data. Therefore, the novel method of data analysis represents an excellent tool for the detection of very sharp phase transitions, e.g., first-order phase transitions.

In this article we will first give a brief summary of the hardware requirements and the conditions under which the method is applicable. We then summarize our previous approach to determine heat capacities using DTA techniques and show how to extract variations in $S$ directly from such DTA data. We finally present selected results from DTA experiments around first-order and second-order phase transitions in the superconductors $YBa_2Cu_3O_7$ and $V_3Si$, respectively.

## II. Hardware and data analysis

### A. Hardware and instrumentation

A typical schematic view of a DTA set-up is shown in Fig. 1. A sample and a reference sample of comparable heat capacity are thermally linked to a large thermal reservoir, the temperature of which can be varied as a function of time. Our measurements are done under high-vacuum conditions and typically at temperatures below 120 K, which allows for an accurate control of the heat links. We assume here that these heat links $k_s$ and $k_r$ that connect the sample (at a temperature $T_s$) and the reference sample (temperature $T_r$), respectively, with the thermal reservoir (temperature $T_b$), obey Fourier's law for the transport of heat, i.e., we are using the fact that the rate of heat flow $P$ is proportional to the temperature difference between the warm and the cold end of each heat link, $P_{r,s} = k_{r,s}(T_b - T_{r,s})$ (in the following, the subscripts $r$ and/or $s$ denote "reference side" and/or "sample side" ). This is certainly fulfilled if the heat links consist of a metallic wire, the thermal conductivity of which does not vary significantly within the temperature interval defined by $T_s$, $T_r$, and $T_b$ (at high enough temperatures, thermal radiation may also have to be taken into account, see Ref. 2, section IIc). This temperature interval is determined by[2]

$$T_b - T_{r,s} = (C_{r,s} + C_A)\dot{T}_b / k_{r,s} = \tau_{r,s}\dot{T}_b, \qquad (1)$$

with the heat capacities $C_r$ and $C_s$ of the reference sample and the sample, respectively, the addenda heat capacity $C_A$, and the time constants $\tau_{r,s} = (C_{r,s} + C_A)/k_{r,s}$. The values for the heat links, $k_s$ and $k_r$, can be estimated from the thermal conductivities, the cross sections and the lengths of the connections used to form the heat links, and may include contributions from wires to the thermometers and from the mechanical suspension of the samples. In typical experiments, we had $(T_b - T_{r,s})/T_b \approx 0.5$ - 3%. The time constants $\tau_{r,s}$ can be directly measured by monitoring the relaxation of $T_s$ and $T_r$ as a function of time following a heat pulse. These $\tau_{r,s}$ data (as functions of temperature) can later be used to calculate the variations in specific heat and entropy. In our experiments we have typical values $\tau_{r,s} \approx 40$ -350 sec and heating rates of the order of $\dot{T}_b \approx 1-20\ \text{mK/sec}$.

The size of the sample, its thermal anchoring to the sample thermometer and the thermometers themselves have to be chosen in a way that the internal thermal equilibrium and response times are much smaller than $\tau_{r,s}$ and smaller than the inverse data-acquisition rate (in our case 3 sec) to guarantee that the $T_s$ and $T_r$ data are not unnecessarily broadened by an internal equilibrium process. As a general trend, a thermal equilibrium in most solids is achieved faster at low temperatures than at room temperature or above, which makes the use of a DTA technique at low temperatures particularly attractive. For our samples (typically a few milligrams) we estimate the respective internal thermal equilibrium times to be of the order of 0,1 sec or less.

As in any thermal experiment, thermometers and wiring should contribute as little as possible to the total heat capacity. We have built various versions of DTA cells using thermocouples and small platinum PT-100 or Cernox 1050 chip resistors. For a quantitative analysis, the corresponding addenda heat capacities $C_A$ and the reference heat capacity $C_r$ have to be known (see below). These quantities, together with $k_{r,s}$, can be reasonably accurately measured using the $T$-dependent $\tau_{r,s}(T)$ data and by performing an additional experiment at constant $T_b$, where the sample and the reference sample are heated with a known constant heating power $P_{r,s}$, thereby raising their respective temperatures by the asymptotic values $(T_{r,s} - T_b) = P_{r,s} / k_{r,s} = P_{r,s}\tau_{r,s} / (C_{r,s} + C_A)$.

The relative sensitivity of both the heat capacity and the entropy data depend crucially on the sensitivity of the sample thermometers and of the way $\Delta T = T_s - T_r$ is meas-ured. Above 50 K we achieved our best results with copper-constantan thermocouples and platinum thermometers (with a typical scattering in $\Delta T$ of the order of $10^{-4}$ K and $10^{-6}$ K, respectively), and at lower temperatures using Cernox thermometers ($10^{-4}$ K). To improve the performance of the resistance thermometers, we applied in some cases a deliberately high measuring current that induced a significant self-heating of the order of 1 K or less. This is not a real problem for the present DTA technique, however, as long as the temperature of the sample is known, the self-heating effect is symmetrical on the sample and the reference sides, respectively, and the quantity $T_{\tau,r} = \tau_r \dot{T}_b \approx T_b - T_r$ that is later used in the analysis is calculated from $\tau_{r,s}$ and $\dot{T}_b$ (note that the thus obtained $T_b - T_{r,s}$ values at large measuring current do not represent the true temperature differences between the sample thermometers and the thermal reservoir during the experiment).

The quantity $\Delta T$ contains the essential information on abrupt changes in $C_s$ and in the associated entropy (see below). It can be measured very accurately in a differential configuration, either as a voltage difference from a differential-thermocouple pair, or as the voltage difference between the centers of two branches of a Wheatstone-like resistance bridge that is configured with two calibrated resistance thermometers (one on the sample and one on the reference side) and two known reference resistors. In our experiments, these DC voltage differences were amplified using a Keithley 1801 preamplifier and measured with a Keithley 2001 voltmeter.

In the following we assume that the quantities $\tau_r(T)$, $C_r(T)$ and $C_A(T)$ have been measured in a separate experiment and are free of any discontinuities. These values have to be determined only once since they are not expected to vary when a sample is exchanged. To eliminate unnecessary sources of scattering to the final result, the $\tau_r(T)$, $C_r(T)$ and $C_A(T)$ data should be smoothed, e.g., by replacing the measured data by continuous functions.

**B. The measurement of heat capacities**

We have described the principle of measuring differences in the heat capacity using this DTA technique in detail in an earlier publication[2]. For a measurement of $C_s(T)$ the thermal reservoir has to be heated or cooled in such a way that the variation $\dot{T}_b$ is known as a function of time. We then have to monitor only the temperature $T_r$ of the reference sample and the temperature difference $\Delta T$, both as functions of time. Like $\tau_r$, $C_r$ and $C_A$, the $\dot{T}_b$ and $T_r$ data will not show any discontinuities as a function of time if the measuring set-up is reasonably constructed and operated, and they can therefore be approximated by smooth continuous functions. This is not possible for $\Delta T$, however, because these data may contain essential information about possible abrupt phase transitions of the sample. Assuming $k_s = k_r$ and equal addenda heat capacities $C_A$ on the sample and the reference side, the heat-capacity difference $\Delta C = C_s - C_r$ at the sample temperature $T_s = T_r + \Delta T$ becomes, to a first approximation,

$$\Delta C \approx -(C_r + C_A)\frac{\Delta T}{T_{\tau,r}}, \tag{2}$$

with $T_{\tau,r} = \tau_r \dot{T}_b$ (note that depending on the sign of $\dot{T}_b$, $T_{\tau,r}$ can be positive or negative, and in the steady state, $T_{\tau,r} = T_b - T_r$). The case of asymmetric heat links, i.e., $k_s \neq k_r$, is discussed in detail in Ref. 2 but it is not considered here. The thus obtained approximate $\Delta C$ data are smeared out on the temperature scale $T_{\tau,s} = \tau_s \dot{T}_b$, and therefore sharp discontinuities in $C_s$ appear to be broadened by this amount. This broadening can be reduced if we use the exact relation[2]

$$\frac{\Delta C}{C_r + C_A} = \frac{\dot{T}_r}{\dot{T}_s}\left(1 - \frac{\Delta T}{T_{\tau,r}}\right) - 1. \tag{3}$$

We can eliminate the time derivative in Eq. (3) using

$$\Delta \dot{T} = \frac{\mathrm{d}\Delta T}{\mathrm{d}T_s}\dot{T}_s \tag{4}$$

and

$$\frac{\dot{T}_r}{\dot{T}_s} = \frac{\dot{T}_s - \Delta \dot{T}}{\dot{T}_s} = 1 - \frac{\Delta \dot{T}}{\dot{T}_s} = 1 - \frac{\mathrm{d}\Delta T}{\mathrm{d}T_s}, \tag{5}$$

and we obtain

$$\frac{\Delta C}{C_r + C_A} = -\frac{\Delta T}{T_{\tau,r}} - \frac{\mathrm{d}\Delta T}{\mathrm{d}T_s} + \frac{\mathrm{d}\Delta T}{\mathrm{d}T_s}\frac{\Delta T}{T_{\tau,r}}. \tag{6}$$

However, the numerical evaluation of $\dot{T}_s$ or $\mathrm{d}\Delta T/\mathrm{d}T_s$ from experimental data again introduces a certain broadening $\delta T$ on the temperature scale that depends on the temperature interval used to calculate this derivative. Moreover, the scatter $\delta\Delta C$ in the $\Delta C$ data as calculated from Eqs. (3) or (6) is inevitably increased by such a procedure. We have shown that the product $\delta\Delta C\delta T$ is a constant that is determined by $C_r$ and by the limiting accuracy with which $\Delta T$ can be measured[2]. In other words, any attempt to increase the accuracy of the $\Delta C$ data in $T$, e.g., by choosing a narrower interval to calculate $\dot{T}_s$ or $\mathrm{d}\Delta T/\mathrm{d}T_s$, leads to an increased scattering in $\Delta C$.

In the next paragraph we describe how we can completely circumvent this problem of instrumental broadening if we consider variations in entropy $S$, rather than variations in the heat capacity $C$.

## C. The measurement of variations in entropy

### a) Measurements with varying temperature

We now consider the difference in the entropy between the sample and the reference sample, $\Delta S = S_s - S_r$ between two sample temperatures $T_1$ and $T_2$,

$$\Delta S(T_2) - \Delta S(T_1) = \int_{T_1}^{T_2} \frac{\Delta C}{T_s} \mathrm{d}T_s \qquad (7)$$

that becomes

$$\Delta S(T_2) - \Delta S(T_1) = \int_{T_1}^{T_2} \frac{C_r + C_A}{T_s} \left( -\frac{\Delta T}{T_{\tau,r}} - \frac{\mathrm{d}\Delta T}{\mathrm{d}T_s} + \frac{\mathrm{d}\Delta T}{\mathrm{d}T_s} \frac{\Delta T}{T_{\tau,r}} \right) \mathrm{d}T_s. \qquad (8)$$

If a sharp phase transition occurs, we restrict on a narrow interval $[T_1, T_2]$ around this transition. We assume here that $C_r + C_A \approx const.$ in this interval, and that the reference sample is in the steady state, i.e., $T_{\tau,r} = \tau_r \dot{T}_b \approx const.$, and therefore

$$\frac{\mathrm{d}T_{\tau,r}}{\mathrm{d}T_s} \approx 0. \qquad (9)$$

It can be shown that if these approximations are not strictly fulfilled, the resulting $\Delta S(T_2) - \Delta S(T_1)$ may at most exhibit a slowly varying systematic error. Partial integration of Eq. (8) and using Eq. (9) gives

$$\frac{\Delta S(T_2) - \Delta S(T_1)}{C_r + C_A} \approx -\int_{T_1}^{T_2} \frac{\Delta T}{T_s T_{\tau,r}} dT_s - \left[ \frac{\Delta T}{T_s} \right]_{T_1}^{T_2} - \int_{T_1}^{T_2} \frac{\Delta T}{T_s^2} dT_s + \left[ \frac{\Delta T^2}{2 T_s T_{\tau,r}} \right]_{T_1}^{T_2} + \int_{T_1}^{T_2} \frac{\Delta T^2}{2 T_s^2 T_{\tau,r}} \mathrm{d}T_s. \quad (10)$$

This result can, in principle, already be used to calculate $\Delta S(T_2) - \Delta S(T_1)$ by numerical integration of experimental data. If the interval $[T_1, T_2]$ is sufficiently narrow (i.e., $|T_2 - T_1| \ll T_s$), we have

$$\left|\int_{T_1}^{T_2} \frac{\Delta T}{T_s^2} \mathrm{d}T_s\right| << \left|\left[\frac{\Delta T}{T_s}\right]_{T_1}^{T_2}\right| \quad \text{and} \quad \left|\int_{T_1}^{T_2} \frac{\Delta T^2}{2T_s^2 T_{\tau,r}} \mathrm{d}T_s\right| << \left|\left[\frac{\Delta T^2}{2T_s T_{\tau,r}}\right]_{T_1}^{T_2}\right|, \tag{11}$$

and we obtain, as a central result of this paper, the simplified expression

$$\frac{\Delta S(T_2) - \Delta S(T_1)}{C_r + C_A} \approx -\int_{T_1}^{T_2} \frac{\Delta T}{T_s T_{\tau,r}} \mathrm{d}T_s - \left[\frac{\Delta T}{T_s} - \frac{\Delta T^2}{2T_s T_{\tau,r}}\right]_{T_1}^{T_2}. \tag{12}$$

The Eqs. (10) and (12) have been derived from the exact result (3), and they therefore do not suffer from any broadening on the temperature scale, i.e., sharp discontinuities in $\Delta S$ will be correctly reproduced in the measured data.

**b) Constant temperature and varying magnetic field**

In some cases it is desirable to keep the temperature of the thermal bath constant while varying the external magnetic field, say from $H_1$ to $H_2$, to probe a possible magnetocaloric effect of the sample. Such experiments are often done under adiabatic conditions (i.e., $k_s = k_r = 0$), where the sample shows a cooling or a heating effect upon the variation of the magnetic field that depends on the temperature and the magnetic-field dependence of the entropy $S(H,T)$. A possible increase in $S$ with increasing magnetic field $H$, for example, typically leads to an adiabatic-cooling effect that balances out the total change in $S$ to zero. Experiments of this type have been reported, for example, on phase transitions in $CeCoIn_5$ at low temperatures[7]. The observed variations in the sample temperature $T_s$ can be very large, however. Therefore, this powerful technique does not allow for isothermal investigations. A sensitive version of an experiment measuring the isothermal magnetocaloric effect using heat-flow sensors has been recently demonstrated to work very well on phase transitions and to study possible irreversible effects, and the relationship of the measured data to corre-sponding magnetic quantities has been discussed in detail[8,9]. To maintain a constant sample temperature in such an experiment, a certain heat flow between the thermal bath and the sample has to supplied, that must be exactly known for the calculation of the thermodynamic quantities of interest. As we have seen in the previous paragraph, the here described DTA technique does not require the knowledge of the amount of heat flowing in the experiment, because it relates the measured temperature differences to the known heat capacities $C_r$ and $C_A$ (see, e.g., Eq. (12)). In typical DTA experiments as we have described above, the $k_s$ and $k_r$ values are chosen in a way that the temperature differences between the thermal bath and the samples are always small compared to the temperature of the experiment (see section II.A). These temperature differences become even smaller when $T_b$ is kept constant and only the magnetic field is varied. Therefore the situation can be regarded as a quasi-isothermal one, and we will show below that we can indeed use such a DTA set-up to measure corresponding variations in entropy, $S(H_2)-S(H_1)$.

At a constant temperature of the thermal reservoir $T_b$ we have $T_r = T_b$, and $\Delta T = T_s - T_b$ (a reference sample is not needed in such an experiment). The change in entropy of the sample and the sample platform is given by the amount of heat $\delta Q$ coming from the thermal reservoir within the time interval d$t$,

$$\mathrm{d}S = \frac{\delta Q}{T_s} = -\frac{\Delta T}{T_s} k_s \mathrm{d}t, \qquad (13)$$

and is equal to

$$\mathrm{d}S = \left(\frac{\partial S}{\partial T_s}\right)_H dT_s + \left(\frac{\partial S}{\partial H}\right)_{T_s} dH = \frac{C_s + C_A}{T_s} dT_s + \left(\frac{\partial S}{\partial H}\right)_{T_s} dH. \qquad (14)$$

Therefore

$$\frac{C_s + C_A}{T_s} \dot{T}_s + \left(\frac{\partial S}{\partial H}\right)_{T_s} \dot{H} = -\frac{\Delta T}{T_s} k_s, \qquad (15)$$

and with $H_{\tau,s} = \dot{H}\,\tau_s$, $\tau_s = (C_s + C_A)/k_s$ and making use of $(\mathrm{d}T_s/\mathrm{d}H)\dot{H} = \dot{T}_s$ we obtain

$$\frac{\left(\frac{\partial S}{\partial H}\right)_{T_s}}{C_s + C_A} = -\frac{\mathrm{d}T_s/\mathrm{d}H}{T_s} - \frac{\Delta T}{T_s H_{\tau,s}} \tag{16}$$

(depending on the sign of $\dot{H}, H_{\tau,s}$ can be positive or negative). If we again assume that a sharp phase transition occurs in the sample, we restrict on an interval [$H_1$,$H_2$] around this transition where $C_s + C_A \approx$ *const.* This requires that any discontinuities in $C_s$ must be small compared to $C_s + C_A$ which has not been necessary to assume in corresponding DTA experiments with varying temperature. Integrating Eq. (16) gives

$$\frac{S(H_2) - S(H_1)}{C_s + C_A} \approx -\ln\frac{T_s(H_2)}{T_s(H_1)} - \int_{H_1}^{H_2} \frac{\Delta T}{T_s H_{\tau,s}}\mathrm{d}H. \tag{17}$$

This result can again be used to calculate $S(H_2) - S(H_1)$ by numerical integration of experimental data.

If we further assume that the variations in $T_s(H)$ upon changing the magnetic field are small compared to the initial value $T_s(H_1)$, we can use

$$\ln\frac{T_s(H_2)}{T_s(H_1)} \approx \frac{T_s(H_2) - T_s(H_1)}{T_s(H_1)} = \frac{\Delta T(H_2) - \Delta T(H_1)}{T_s(H_1)}, \tag{18}$$

and

$$\frac{S(H_2) - S(H_1)}{C_s + C_A} \approx -\left[\frac{\Delta T}{T_s}\right]_{H_1}^{H_2} - \int_{H_1}^{H_2} \frac{\Delta T}{T_s H_{\tau,s}}\mathrm{d}H. \tag{19}$$

We may identify these changes in $S$ with changes in the sample entropy $S_s$ alone if the sample platform does not show any magnetocaloric effect.

Is worth mentioning that we can derive the thermodynamic quantity $(\partial M/\partial T)_H$ from experimental data obtained using Eq. (19) by numerically differentiating correspond-ing isothermal entropy $\Delta S(H)$ data and using the Maxwell relation $(\partial S/\partial H)_T = (\partial M/\partial T)_H$.

## III. First and second-order phase transitions in superconductors

In Figs. 2 and 3 we show, as an example, variations in entropy that we have measured on a single crystal of the high-temperature type-II superconductor $YBa_2Cu_3O_7$ with a sample mass $m$ = 3.3 mg. The experimentally determined magnetic phase diagram of this sample[6] is shown in the inset of Fig. 2b. In the mixed state the vortex lattice undergoes a fairly sharp first-order phase transition from a solid to a so-called vortex-fluid state at a phase boundary $H_m(T)$. In Fig 2a we present the result of a DTA measurement around this transition with a fixed magnetic field $\mu_0 H$ = 5 T and varying temperature $T$. For this experiment, we used PT-100 thermometers with a large excitation current $I$ = 3.5 mA, at a heating rate $\dot{T}_b \approx +17$ mK/sec and $\tau_r \approx 40$ sec. The $\Delta S$ data shown in Fig. 2a have been obtained from the $\Delta T$ raw data plotted in the same figure and using Eq. (12). For clarity, linear backgrounds in $\Delta S$ and $\Delta T$ that fit the respective data right below the phase transition have been subtracted. From such measurements, the total heat capacity (see Fig. 2b) can be also obtained as discussed above and in more detail in Ref. 2.

In Fig. 3 we present the result of corresponding measurements at fixed temperature $T$ = 83.1 K and with a varying magnetic field $H$. These experiments have been done using a pair of copper-constantan thermocouples in a differential configuration, for both increasing and decreasing magnetic field with $\mu_0 \dot{H} \approx \pm 1.67$ mT/sec and with $\tau_s \approx 90$ sec. The $\Delta T$ raw data plotted in Fig. 3a have been treated according to Eq. (19) to obtain the resulting $\Delta S(H)$ shown in Fig 3b. Again, linear backgrounds in $\Delta S$ and $\Delta T$ that fit the respective data right below (measurements with $H$ increasing) or above the phase transition ($H$ decreasing) have been subtracted for clarity reasons.

In both types of DTA measurements we observe a distinct step in entropy ($\Delta S \approx 1.4$ mJ/moleK$^2$) at the phase-transition line $H_m(T)$, with a Gaussian half-width of 90 mK on the temperature scale and 50 mT on the magnetic-field scale, respectively, indicated by the shaded transition regions in Figs. 2a and 3a. It is important to note here that the "natural" instrumental temperature and magnetic-field scales that determine the broadening of the $\Delta T$ raw data are $T_{\tau,r} = \tau_r \dot{T}_b$ and $H_{\tau,s} = \dot{H}\tau_s$, respectively. The $\Delta T$ data in Figs. 2a and 3a are indeed smeared out outside the transition region after the transition is completed. The measured half widths in entropy are significantly smaller than these values ($T_{\tau,r} = 680$ mK and $H_{\tau,s} = 150$ mT), and represent the intrinsic values for this phase transition in the considered sample.

In Fig. 4 we finally show a DTA measurement on a single crystal of $V_3Si$ with a sample mass $m$ = 11 mg around its second-order phase transition to superconductivity at $T_c \approx 16.7$ K in zero magnetic field. For this experiment, we used Cernox 1050 thermometers with a small excitation current $I$ = 2 μA at a heating rate $\dot{T}_b \approx +2.3$ mK/sec and $\tau_s \approx 320$ sec. The $\Delta S$ data shown in Fig. 4a have again been obtained from the $\Delta T$ raw data plotted in the same figure and using Eq. (12) (again, a background in $\Delta S$ has been subtracted for clarity reasons). In Fig. 4b we show the resulting specific heat $C/T$, here obtained from Eq. (6) and by numerically calculating the derivative $\mathrm{d}\Delta T/\mathrm{d}T_s$. As in the above example, the total intrinsic transition width in $\Delta S$ (≈ 130 mK, as indicated by the shaded region) is much smaller than the instrumental broadening of the $\Delta T$ raw data, $T_{\tau,r} = 740$ mK (see Fig. 4a). According to the discussion in sections I and II of this paper, the instrumental resolution in $T$ or $H$ in such experi-ments is indeed limited only by the internal thermal equilibrium time of the sample (including the sample platform and its suspension), and by the response time of the used thermometers.

**Acknowledgements**

This work was supported by the Schweizerische Nationalfonds zur Förderung der Wissenschaftlichen Forschung, Grant. No. 20-111653.

**Figure captions**

*Fig. 1:* Schematic view of a standard DTA configuration. The sample and a reference sample with the heat capacities $C_s$ (at a temperature $T_s$) and $C_r$ (temperature $T_r$) are connected via the heat links $k_s$ and $k_r$ with a thermal reservoir (temperature $T_b$). The difference $\Delta T = T_s - T_r$ and $T_r$ are monitored during the variation of $T_b$ or the external magnetic field $H$ as a function of time. With known $C_r$ and addenda heat capacity $C_A$, entropy changes can be calculated using Eqs. (12) or (19) as discussed in the text.

*Fig. 2: (a)* Results of a DTA experiment on a $YBa_2Cu_3O_7$ single crystal in a fixed magnetic field and varying $T_b$. For clarity, linear backgrounds in $\Delta S$ and $\Delta T$ that fit the respective data right below the phase-transition region (grey-shaded region) have been subtracted to obtain the variations $\delta\Delta S$ (left scale) and $\delta\Delta T$ (right scale), respectively (see text). $T_{\tau,r}$ illustrates the instrumental broadening of the $\Delta T$ raw data. *(b)* Total heat capacity $C/T$ with the first-order phase transition at $T \approx 83.0$ K. The inset shows the magnetic phase diagram of $YBa_2Cu_3O_7$. Arrows indicate the direction in which the DTA runs have been made to obtain the data plotted in Figs. 2 and 3.

*Fig. 3: (a)* Results of a DTA experiment on $YBa_2Cu_3O_7$ at fixed temperature and varying magnetic field. Arrows indicate the direction of the field scans. The phase-transition region is indicated by the grey-shaded area. $H_{\tau,s}$ indicates the instrumental broadening of the $\Delta T$ raw data. *(b)* Variations in entropy obtained form the $\Delta T$ raw data in Fig. 3a. Linear backgrounds in both $\Delta S$ and $\Delta T$ that fit the respective data around the phase transition have been subtracted for clarity reasons (see text).

*Fig. 4: (a)* Results of a DTA experiment (entropy $\Delta S$ and $\Delta T$ raw data) on a $V_3Si$ single crystal in zero magnetic field and with varying $T_b$. A background in $\Delta S$ that fits the respective data right above the second-order phase-transition region to superconductivity (grey-shaded region) has been subtracted to obtain the variation $\delta\Delta S$. $T_{\tau,r}$ shows the instrumental broadening of the $\Delta T$ raw data. *(b)* Total heat capacity $C/T$ of $V_3Si$ around $T_c \approx 16.7$ K as obtained from the data in Fig. 4a.

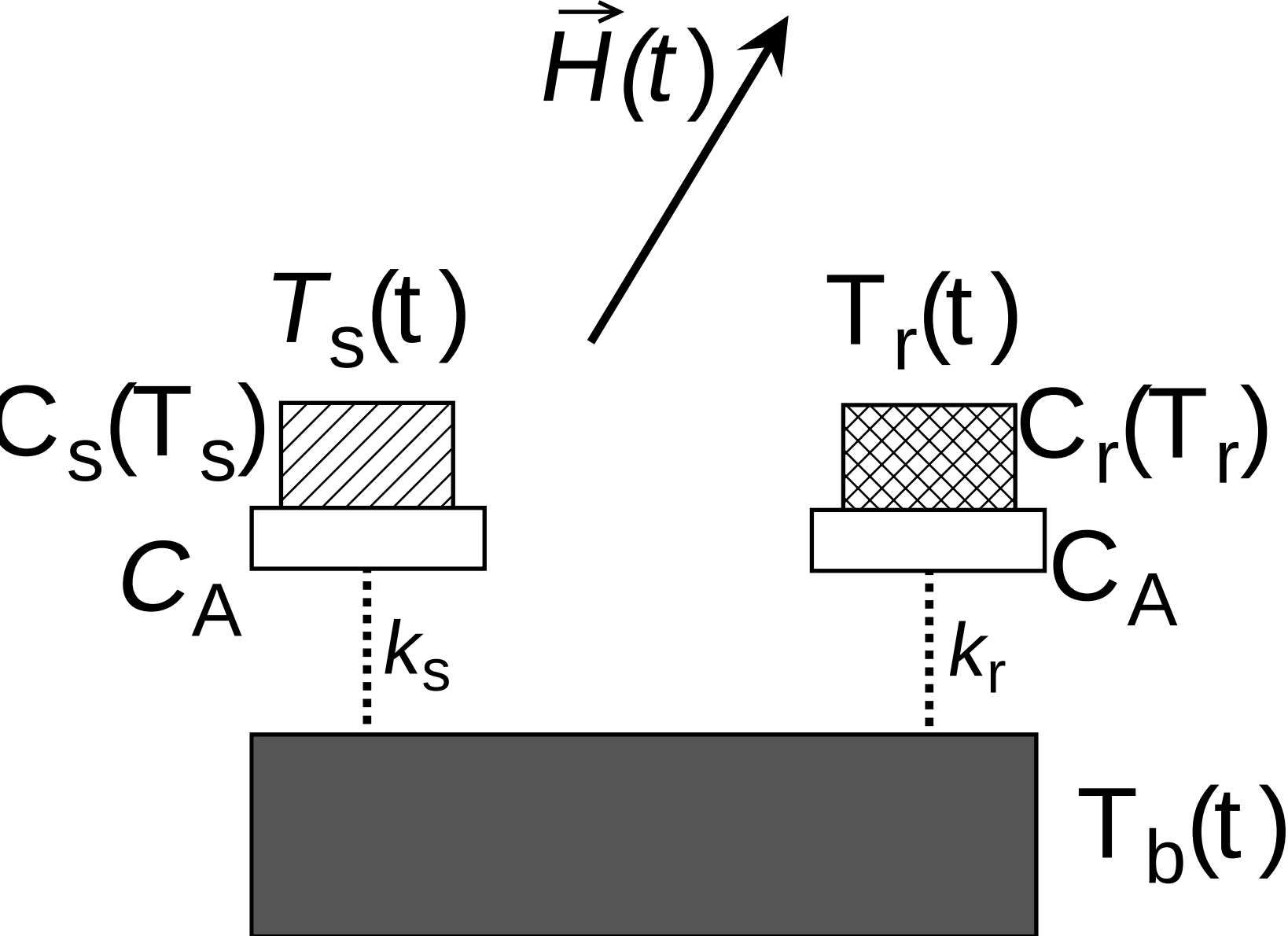

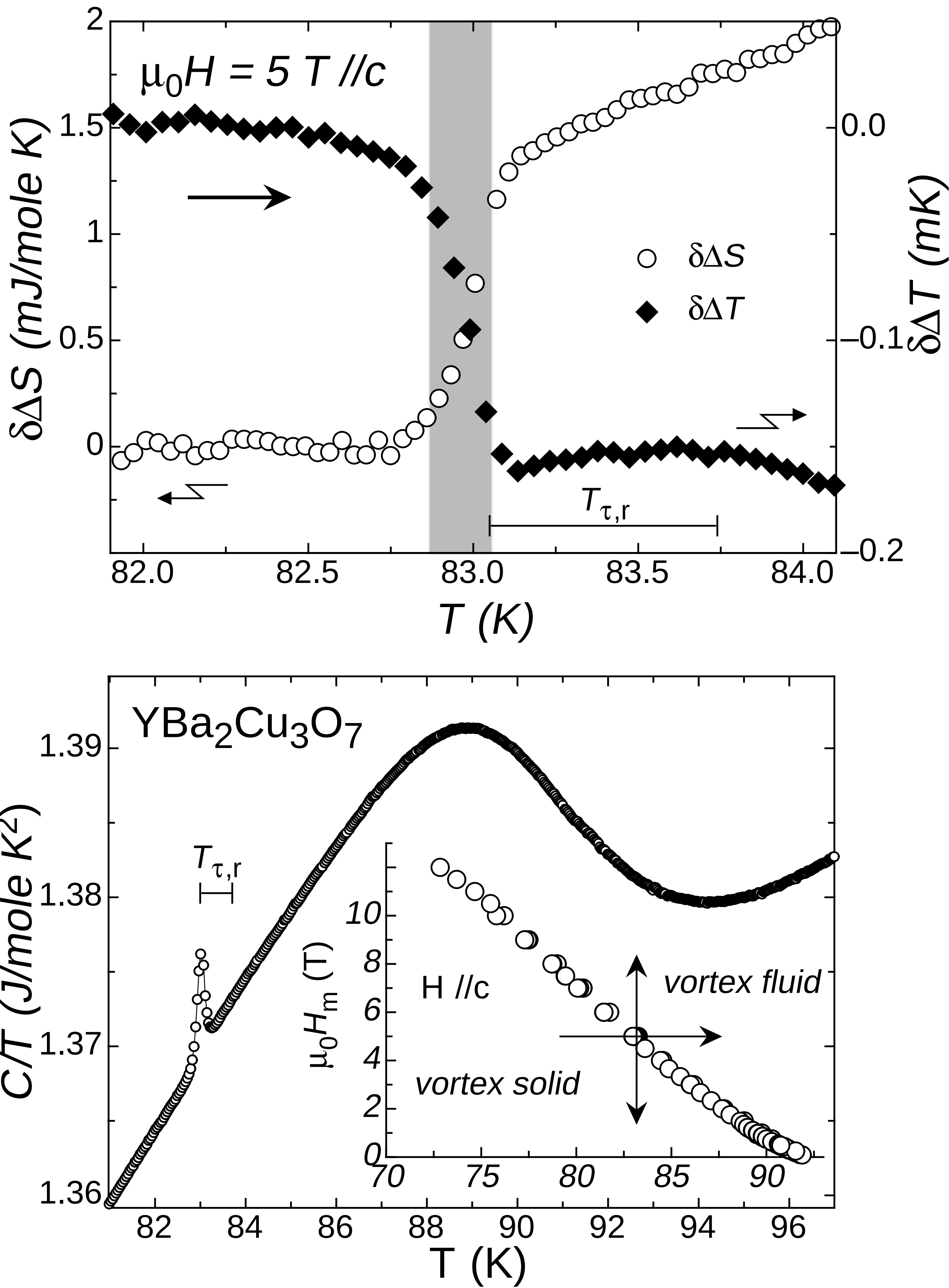

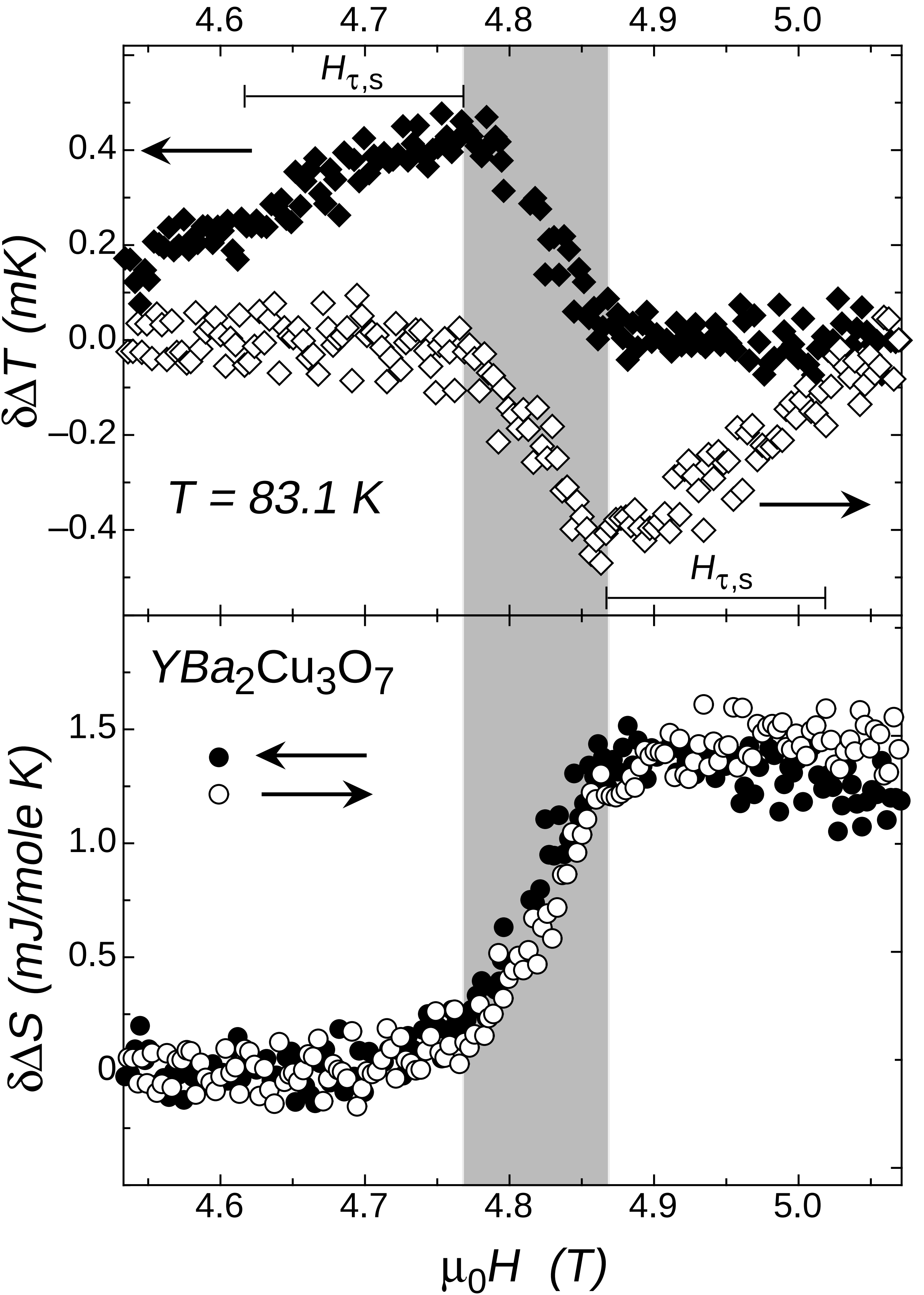

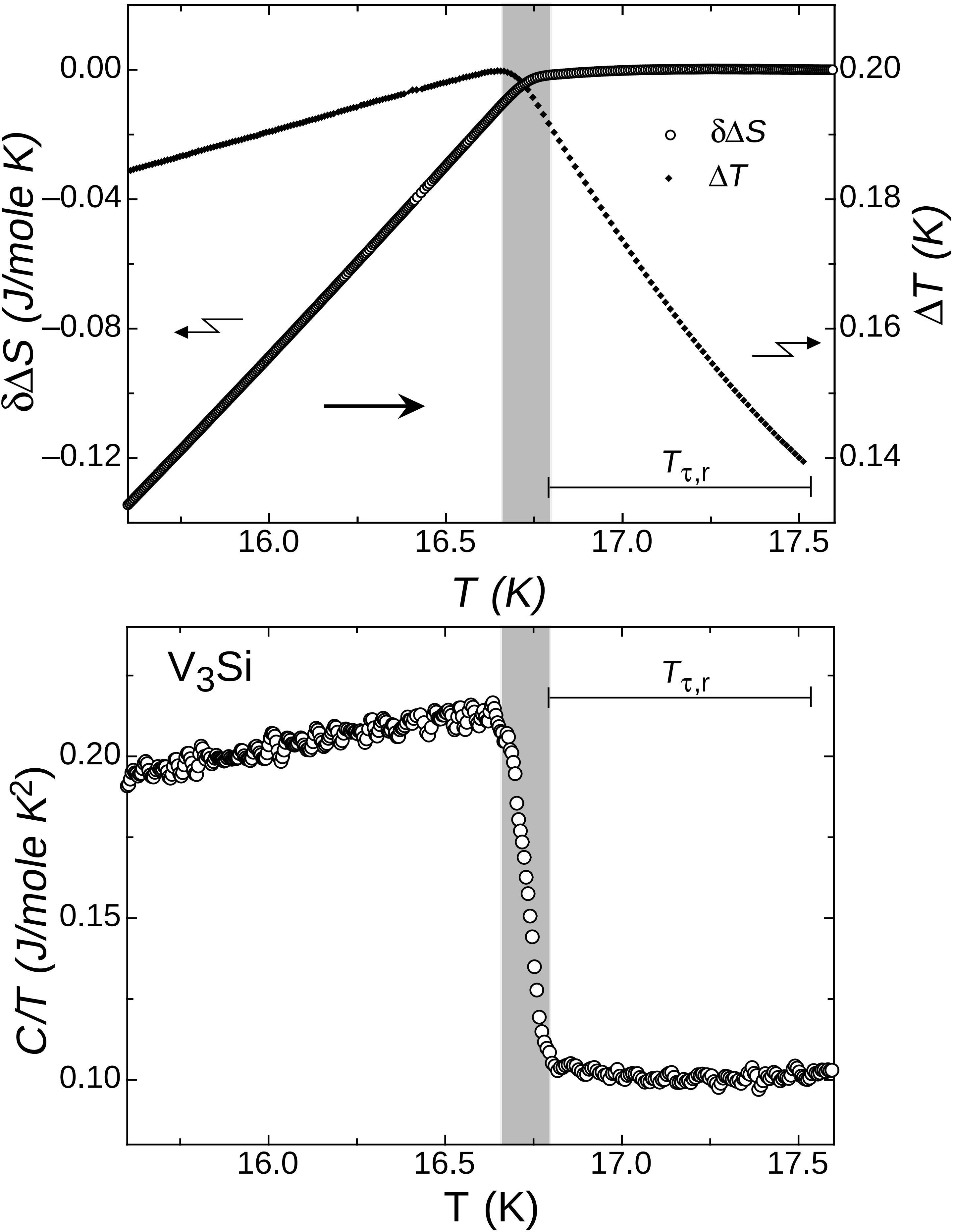